\documentclass[structabstract]{aa}
\pdfoutput=1 
\usepackage{graphicx}
\begin{document}
\authorrunning{I.F. Mirabel et al.}
\titlerunning{Stellar black holes at the dawn of the universe}

   \title{Stellar black holes at the dawn of the universe}

   \author{I.F. Mirabel \inst{1,2}, M. Dijkstra \inst{3,5}, P. Laurent \inst{1,4}, A. Loeb \inst{3}, J.R. Pritchard \inst{3}}

   \institute{CEA-Saclay, IRFU/DSM/Service d'Astrophysique. 91191 Gif/Yvette. France\\
              \email{felix.mirabel@cea.fr}\\
         \and
    Instituto de Astronom\'ia y F\'isica del Espacio. cc 67, suc. 28. (C1428) Buenos Aires. Argentina \\
         \and
    Harvard-Smithsonian Center for Astrophysics, 60 Garden Street, Cambridge, MA 02138. USA \\
         \and
    Laboratoire APC, 10 rue Alice Domont et L\'eonie Duquet, 75205, Paris, France \\
\and
    Max Planck Institute f\"{u}r Astrophysik, Karl-Schwarzschild-Str. 1. 85741 Garching, Germany\\
             }

   \date{Received 21/12/2010; Accepted 27/01/2011  }


  \abstract{It is well established that between 380000 and 1 billion years after the Big Bang the Inter Galactic Medium (IGM) underwent a ``phase transformation" from cold and fully neutral to warm
($\approx 10^4$ K) and ionized. Whether this phase transformation
was fully driven and completed by photoionization by young hot stars is a question of topical interest in cosmology. }{We propose here that besides the ultraviolet radiation from massive stars, feedback from accreting black holes in high-mass
 X-ray binaries (BH-HMXBs) was an additional, important source of heating and reionization of the IGM in regions of low
 gas density at large distances from star-forming galaxies.}{We use current theoretical models on the formation and evolution of primitive massive stars of low metallicity, and the observations of compact stellar remnants in the near and distant universe, to infer that a significant fraction of the
 first generations of massive stars end up as BH-HMXBs. }{The total number of energetic ionizing photons from an accreting stellar black hole in an HMXB is comparable to the total number of ionizing photons of its progenitor star. However, the  X-ray photons emitted by the accreting black hole are capable of producing several secondary ionizations and the ionizing power of the resulting black hole could be greater than that of its progenitor.  
 Feedback by the large populations of BH-HMXBs heats the IGM to
temperatures of $\approx 10^4$ K and maintains it ionized on large
distance scales.}{BH-HMXBs determine the early thermal history of the universe and mantain it as ionized over large volumes of space in regions of low density. This has a direct  impact on the properties of the faintest galaxies at high redshifts, the smallest dwarf galaxies in the local universe, and on the existing and future surveys at radio wavelengths of atomic hydrogen in the early universe.}


   \maketitle
%

\section{Introduction}

The so-called ``dark ages" of the universe started $\approx 380000$
years after the Big Bang as matter cooled down and
 space became filled with neutral hydrogen. This phase of the universe lasted up to about a billion years, when the complex process of reionization
 of the IGM was completed (\cite{Loeb10}). It is currently believed that most of the reionization was caused by the ultraviolet radiation from massive stars formed in the first generations of galaxies. However, it is uncertain what fraction of ionizing 
ultraviolet photons could escape from primitive galaxies to produce and maintain the ionization far from galaxies in low density regions  of the IGM.  Recent observations with the Hubble Space Telescope suggest that the rest-frame ultraviolet radiation from the most distant galaxies detected so far at the heart of the dark ages is not enough to heat and ionize the IGM over large volumes of space. To solve this apparent ``photon-starved problem" (\cite{Bouwens10}), it has been suggested by Lehnert et al. (2010) that a fainter population of galaxies below the present detection limit could contribute significantly to the reionization. 

X-rays from accreting black holes have a longer mean free path than the
ultraviolet photons from massive stars. In this context Madau et al. (2004) and Ricotti \& Ostriker (2004) have suggested that a smaller fraction of ionizing photons were provided by primordial black holes of
intermediate mass (``miniquasars", at z $> 10$) accreting by
Bondi-Hoyle from the surrounding gas. However, feedback
from Bondi-Hoyle accretion to solitary black holes significantly
suppresses both any further inflow (\cite{Alvarez09}; \cite{Milo09}) and the
consequent injection of radiation and high-energy particles to the
surrounding medium. 

We propose that  BH-HMXBs at z $\geq$ 6, namely, the fossils of massive stars are an important -so far overlooked- agent in the complex process of the reionization of the universe. In the context of the current models
on the formation (\cite{Krumholz09}; \cite{Turk09}; \cite{Stacy10})
and collapse (\cite{Heger03}; \cite{Meynet05}; \cite{Georgy09}; \cite{Linden10}) of primordial stars,
an attractive and realistic alternative to the hypothesis of
quasi-radial, Bondi-like accretion on solitary black holes of
intermediate mass (``miniquasars") is accretion on stellar black
holes  from high-mass stars in binary
systems, namely, ``microquasars" (\cite{Mirabel99}). As shown below, the
formation rate of BH-HMXBs must have been very large in the young
Universe, playing an important role in the thermal history of the
IGM, and a complementary role to that of
their progenitor stars in the re-ionization process of the IGM over
large volumes of space. It is this scenario that we investigate here.

\section{Cosmic evolution of BH-HMXBs: A prediction from current theoretical models }
Recent hydrodynamic simulations of the formation of the first
generations of stars show that a substantial fraction of stars in
primordial galaxies form as binaries with typical masses of tens of
solar masses (\cite{Krumholz09}; \cite{Turk09}; \cite{Stacy10}).
Models of single stars with very low metal content and initial masses of a few tens
of solar masses show that they  collapse directly with no energetic natal kicks, and
end as black holes  (\cite{Heger03}; \cite{Meynet05};
\cite{Georgy09}). 

On the other hand, a recent model of binary evolution of massive stars 
by Linden et al. (2010) show that the number of HMXBs and ULXs, their time evolution, and orbital 
period distribution, are strongly metallicity dependent. Linden et al. (2010) find that ULXs formed  in a typical starburst 
of 10$^6$ $M_{\sun}$  with Z = 0.02 Z$_{\sun}$ outnumber ULXs formed with Z = Z$_{\sun}$ by a factor of 5, and after 10 Myr by almost three orders of magnitude. Besides, at Z = 0.02 Z$_{\sun}$, among the ULX population, $>95\%$  of the compact objects are black holes formed by direct collapse and therefore after black hole formation it remains gravitationally bound to a companion, donor star. Most of the orbital periods at Z = 0.02 Z$_{\sun}$  are less than 3 days and accretion is by Roche lobe overflow which creates very luminous and persistent  BH-HMXBs. Probably, this trend continues for starbursts with the metallicities (Z $\leq 0.02 Z_{\sun}$) of the reionization era.
       
These models imply that the majority
of the first generations of high mass stellar binaries 
remain gravitationally bound after the formation of black holes.
Massive stellar binaries can thus become BH-HMXB microquasars, which are sources
of UV photons, X-rays, massive winds, and relativistic jets  (\cite{Mirabel99}). 
Therefore, in the context of the models of massive stellar evolution and
the cosmic evolution of metallicity it is expected that: \textit{1)
the fraction of black holes to
neutron stars and 2) the fraction of black hole binaries to solitary
black holes, should increase with redshift. That is, the rate of
formation of bright BH-HMXBs was likely much larger in the early Universe
than at present.}

\section{Formation rate of stellar black holes as a function of  metallicity: Observations }
The cosmic evolution of BH-HMXBs inferred from theoretical models is consistent with the following
observational studies of stellar black holes and neutron stars in the near and distant universe:
   \begin{enumerate}
      \item  The mass of black holes in HMXBs seems to be a decreasing function of the host galaxy metallicity (\cite{Crowther10} and references therein). 
The black holes in the binaries M 33 X-7, NGC 300 X-1, and IC10 X-1 
are in low-metallicity galaxies and have masses -determined dynamically- in the range of 16 to 30 solar masses, which are higher than the mass of any known
      stellar compact source in the Milky Way and Andromeda galaxies, which have higher metallicities.{ However,  while the model by Linden et al. (2010) supports 
the formation of HMXBs and  ULXs in low-metallicity environments, they conclude that it is difficult to create very massive black holes through common envelope phases, 
since this tends to strip a high fraction of the primary envelope. Given the low number statistics of the known dynamic masses of 
black holes in HMXBs, it is possible that the relatively high masses of 16 to 30 solar masses come from the selection of the brightest sources, namely, those that are at the tip of the iceberg.}           
      \item  It is believed that the majority of ultraluminous X-ray sources (ULXs) found in external galaxies are HMXBs that
      contain black holes accreting at Super-Eddington rates (\cite{Gladstone09}). In fact, 
      the occurrence rate per unit galaxy mass of ULXs observed in nearby galaxies is a decreasing function of the galaxy mass - hence of the metallicity - of the host galaxy (\cite{Zamperi09}).
      \item The space kinematics of Galactic X-ray binaries that contain black holes with more than ten solar masses provides
      evidence of black hole formation by implosion, with no large kicks due to energetic supernovae (\cite{Mirabel03}; \cite{Mirabel10}).
     \item  Observations now support the notion that massive stars with high metal content may end as neutron stars instead of  black holes (\cite{Meynet05}; \cite{Georgy09}; \cite{Linden10}). 
Recently formed neutron stars observed as soft gamma ray repeaters and
     anomalous X-ray pulsars are found in young clusters of large metal content that contain stars with masses of 40-50
     solar masses (\cite{Figer05}; \cite{Muno06}).
     \item It is believed that the majority of gamma ray bursts of long duration (LGRBs) mark the formation of black holes by the collapse of massive stars.
     Although a fraction of dark LGRBs may require local extinction columns of Av $>1$ mag, the majority of their hosts are faint,
     irregular galaxies with global limited chemical evolution  (\cite{LeFloch03}; \cite{Fruchter06}; \cite{Han10}; \cite{Levesque10}).
     The properties of GRB 090423 at z = 8.1 are similar to those of GRBs observed at low/intermediate redshifts (\cite{Salvaterra09}),
     suggesting that the mechanisms and progenitors that gave rise to this burst are not too different from those producing GRBs with
     identified hosts.
     \item There is increasing evidence for an enhanced LGRB rate at z $> 3$ 
(\cite{Daigne06}; \cite{Kistler08}; \cite{Wanderman10}; \cite{Qin10}), as expected from the increase in the specific star
     formation rate (SFR) with decreasing metallicity (\cite{Mannucci10}).
   \end{enumerate}

\section{Ionizing power of a stellar black hole in an HMXB relative to its progenitor star}
In the following we compute the number of ionizing soft X-rays and UV photons from the accretion disk of a BH-HMXB, and then compare its ionization power with that of its progenitor massive star. To this end we assume that the black hole of mass M$_{\rm{BH}}$ is accreting at a fraction f$_{\rm{edd}}$ of its Eddington luminosity for a time t$_{\rm{acc}}$. Then the ratio of the total number of ionizing photons emitted by the accreting black hole to that emitted by the progenitor star is given by
 
 \begin{equation}
\begin{array}{l}
      \frac{\mathrm{N_{\gamma,BH}}}{\mathrm{N_{\gamma,*}}} =
0.6
\left(\frac{\mathrm{M_{BH}}}{\mathrm{M_*}}\right)
\left(\frac{\mathrm{f_{Edd}}}{\mathrm{0.1}}\right)
\left(\frac{\mathrm{t_{acc}}}{\mathrm{20 Myr}}\right) 
\left(\frac{\mathrm{f_{esc,BH}}}{\mathrm{1.0}}\right) 
\left(\frac{\mathrm{N_{phot}}}{\mathrm{64000}}\right)^{-1}\\
\hspace{4.7cm}
\left(\frac{\mathrm{\langle E_{\gamma}}\rangle}{\mathrm{keV} }\right)^{-1}
\left(\frac{\mathrm{f_{esc,*}}}{\mathrm{0.1}}\right)^{-1}
\,,
\end{array}
 \end{equation}

\noindent where  $N_{phot}$ denotes the number of ionizing photons emitted per hydrogen nucleus involved in star formation, $\langle E_{\gamma} \rangle$  denotes the mean photon energy of the radiation emitted by the accreting black hole, and f$_{\rm{esc,*}}$  (f$_{\rm{esc,BH}}$) denotes the fraction of ionizing photons emitted by the star (accreting black hole) that escape from the galaxy and contribute to the heating and reionization of the IGM.

We have substituted reasonable numbers for each of the parameters: N$_{\rm{phot}}$=64000 corresponds to metal-free forming stars with a top-heavy Initial Mass Function (IMF). For a normal stellar population this number would be 16 times lower (\cite{Schaerer03}). The escape fraction of ionizing photons emitted by stars was taken to be f$_{\rm{esc,*}}=0.1$, which is consistent with the mean observed escape fraction of ionizing photons from star-forming galaxies (\cite{Shapley06}) at z=2-3. 

HMXBs are expected to inject photons into the IGM at a rate close to Eddington (and possibly super Eddington), and our choice  f$_{\rm{edd}}=0.1$ is likely to be conservative  (see footnote \footnotemark[1]). The accretion lasts t$_{\rm{acc}} \approx 20$ Myrs, a mean lifetime of donor stars of $M_*=10-30 M_{\sun}$ (\cite{Turk09}; \cite{Stacy10}). The mean photon energy of the radiation emitted by the accreting source depends on the assumed spectral shape, but ${\langle E_{\gamma}}\rangle=1$ keV appears -within a factor of a few- of values derived for both the thermal and power-law components of the black hole spectra.  Finally, f$_{\rm{esc,BH}}=1.0$ is not well constrained. We took f$_{\rm{esc,BH}}$  to be larger than f$_{\rm{esc,*}}$  simply because energetic photons can propagate much more easily through HI column densities in the range $10^{17}-10^{20} \rm{cm}^{-2}$. For reasonable choices for each of these model parameters, we find that  \textit{an accreting black hole in a high-mass binary emits a total number of ionizing photons that is comparable to  its progenitor star}.

However, it should be kept in mind that the ionizing photons emitted by the accreting black hole are more energetic than those from the progenitor star and capable of ionizing more than one hydrogen atom. In a fully neutral medium, the number of secondary ionizations $\rm{N}_{sec} = 25(E_{\gamma}/1 \rm{keV})$, where $E_{\gamma}$ is the photon energy (\cite{Shull85}).  Therefore, the ionizing power of 
the resulting black hole could be greater than that of its progenitor.

\section{BH-HMXBs and massive star formation rates in the epoch of reionization}
The progenitor star of the black hole should have formed in a molecular cloud, which in turn may have formed more stars, each of which contributed to the total number of ionizing photons emitted by stars. Therefore, in the following we look at the emission of ionizing radiation from a star-forming region as a whole.

Observations of galaxies in the local universe show that their X-ray luminosity in the energy range 2-10 keV correlates strongly with the rate at which they are forming stars (\cite{Grim03}). This local correlation states that the X-ray luminosity (in erg s$^{-1}$) scales with the SFR in M$_{\sun}$ yr$^{-1}$, as L$_{2-10} = 7 ~ 10^{39}$ SFR erg s$^{-1}$. When modeling the impact of X-ray emission from galaxies at very high redshift, the following more general correlation is used (\cite{Furlanetto06a}):

\begin{equation}
 L_{2-10} = \rm{f_X} \times 3.5~ 10^{40}  \times \rm{SFR ~~erg~ s^{-1}}.
\end{equation}

Here, the parameter f$_X$  accounts for the likely case that the normalization of the observed correlation depends on redshift. Observations indicate that f$_X = 0.2$ for local galaxies. This important parameter f$_X$ depends on several physical processes (all of which are expected to change with redshift). To illustrate this quantitatively, we express the X-ray luminosity of a star-forming galaxy in more fundamental quantities as

\begin{equation}
 \begin{array}{l}
 L_{2-10}= \\
    \rm{f_{2-10}} \times \frac{dM_{BH}}{dt} \times t_{acc} \times f_{bin} \times f_{Edd} \times 1.5 10^{38} \rm{erg~s^{-1} M_{\sun}^{-1}}
\end{array}
\end{equation}

or

\begin{equation}
 \begin{array}{l}
 L_{2-10}= \rm{f_{2-10}} \times f_{BH} \times {SFR} \times t_{acc} \\
\hspace{2.5cm}
\times \rm{f_{bin}} \times f_{Edd} \times 1.5~10^{38} \rm{erg~s^{-1} M_{\sun}^{-1}}
\end{array}
\end{equation}

Most of the parameters in this equation were introduced earlier.  The fraction f$_{2-10}$ denotes the fraction of the total luminosity that emerges in the 2-10 keV band. For the power law component of the spectra we expect f$_{2-10} = 0.1 - 0.5$ (equal flux per logarithmic bin of energy or steeper), and we conservatively adopt f$_{2-10} = 0.1$. We also introduced the parameter f$_{BH}$, which relates black hole formation rate dM$_{BH}$/dt to SFR. This fraction can be computed for a given initial IMF under the assumption that every star above some critical mass M$_{*,\rm{crit}}$ ends up as black hole. As argued previously, due to mass lost by metallicity-dependent stellar winds even massive stars may end up as neutron stars instead of directly as black holes. However, in the evolution of a close massive binary of low metallicity, due to mass transfer and common envelope phase the primary could  lose mass and end its life as a neutron star rather than a black hole, leading to suppression of black hole formation.  Linden et al. (private communication) discussed the importance and frequency of such scenario finding that the transition from neutron star to black hole dominated HMXBs is very sharp and metallicity dependent, and that for $Z \leq 0.02 Z_{\sun}$  all primary stars  with $M \geq 20 M_{\sun}$  in HMBs end as black holes in HMXBs. Therefore we assume here that mass transfer in close binaries has little impact on the mass range of stars becoming black holes. For the conservative choices of a Salpeter IMF in the range M$_{\rm{low}} = 0.1 M_{\sun}$, M$_{up} = 100 M_{\sun}$, and M$_{*,\rm{crit}} = 25 M_{\sun}$ we find f$_{BH} = 0.03$. 
 Finally, we expect black holes in binaries to be very efficient accreters, while isolated black holes are not. The parameter f$_{bin}$ denotes the mass fraction of accreting black holes in binaries. The total mass fraction of massive stars in binaries is close to unity, and for simplicity we assume that this total mass fraction is $50~ \%$. Since BH-HMXBs are persistent sources of radiation and jets for our model we assume f$_{bin} = 0.5$.

Both Eqs. (2) and (4) depend linearly on SFR, and we can express the parameter f$_X$ as a combination of physical parameters:

\begin{equation}
 f_{X} =  \frac {\rm{f_{2-10}} \times f_{BH} \times t_{acc} \times f_{bin} \times f_{Edd} \times 1.5~10^{38}} {3.5~10^{40}} \end{equation}

or

\begin{equation}
 f_{X}   =
4.0
\left(\frac{\mathrm{f_{2-10}}}{\mathrm{0.1}}\right)
\left(\frac{\mathrm{f_{BH}}}{\mathrm{0.01}}\right)
\left(\frac{\mathrm{f_{Edd}}}{\mathrm{0.1}}\right)
\left(\frac{\mathrm{f_{bin}}}{\mathrm{0.5}}\right)
\left(\frac{\mathrm{t_{acc}}}{\mathrm{20 ~Myr}}\right)
\end{equation}

 Linden et al. (2010) estimate that at ULX luminosities the number of Z = 0.02 Z$_{\sun}$ sources should outnumber the Z = Z$_{\sun}$ HMXBs by a factor of 5. Therefore, in our fiducial conservative model for primordial starbursts f$_X$ is at least one order of magnitude higher than the locally observed value (\cite{Grim03}). There are several reasons why we expect f$_X$ to be much higher in the young Universe. (1) Microquasars can have spectra that are harder and f$_{2-10}$ could be in some cases as high as f$_{2-10} = 0.5$; and (2) f$_{BH}$ is higher in low metallicity environments,  i.e., the formation of black holes in metal-enriched environments of galaxies in the local universe is likely to be strongly suppressed compared to more pristine environments. (3) Stars in the young Universe wre very likely formed with an IMF that was more top-heavy. 
This evolution in the IMF alone could
 boost f$_{BH}$ to much higher values than the fiducial assumed value in Eq. (4). Finaly, the IGM at z = 10 would  essentially be transparent to the hard X-ray photons ($> 1$ keV),
 and following Dijkstra, Haiman \& Loeb (2004) we explicitly verified that these models are consistent with the observed unresolved soft X-ray background.

\footnotetext[1]{To estimate the ionizing power of BH-HMXBs in the early universe we use the observed spectra of the Galactic black hole binary Cygnus X-1,  which has a blue supergiant companion, as template for moderate accretion rates. In this system, accretion is persistent and composed mainly by the donor stellar wind. Its X-ray spectrum is characterized by two components: a UV-soft X-ray bump due to thermal emission from the accretion disk with a typical temperature of $\approx 7$ eV, and a non-thermal power-law component of hard X-rays that result from Compton up scattering of thermal accretion disk and synchrotron photons by a hot coronal plasma and/or jet.  In addition to X-rays, these sources can indeed produce jets and winds of accelerated particles, which in turn may heat and ionize the surrounding medium. \\
On the other hand, for higher accretion rates one can use the ULXs observed in the local universe as templates of BH-HMXBs in the early universe. ULXs often have spectra that resemble the very high (super-soft) state observed in some Galactic black hole binaries (e.g. GRS 1915+105). The majority of ULXs exhibit a complex curvature, which can be modeled by a cool disk component, together with a power law that breaks above 3 keV, probably due to a cool, optically thick corona produced by super-Eddington accretion flows (\cite{Gladstone09}).  Examples of steady super-Eddington sources are also SS 433 in the Milky Way, which is blowing the nebula W50  laterally, and the microquasar that is inflating the nebula S26 in the galaxy NGC 7793 (\cite{Pakull10}). SS 433 injects more than $10^{39}$ erg s$^{-1}$  in the interstellar medium and the microquasar in S26 more than $10^{40}$ erg s$^{-1}$. The overall energy injected by these microquasars during their whole lifetime can be  more than $10^{54} erg$, which is orders of magnitude more than the photonic and baryonic energy from a typical core collapse supernova.}

\section{Stellar black holes and the thermal history of the IGM}
\begin{figure}
   \centering
   \includegraphics[width=8.cm]{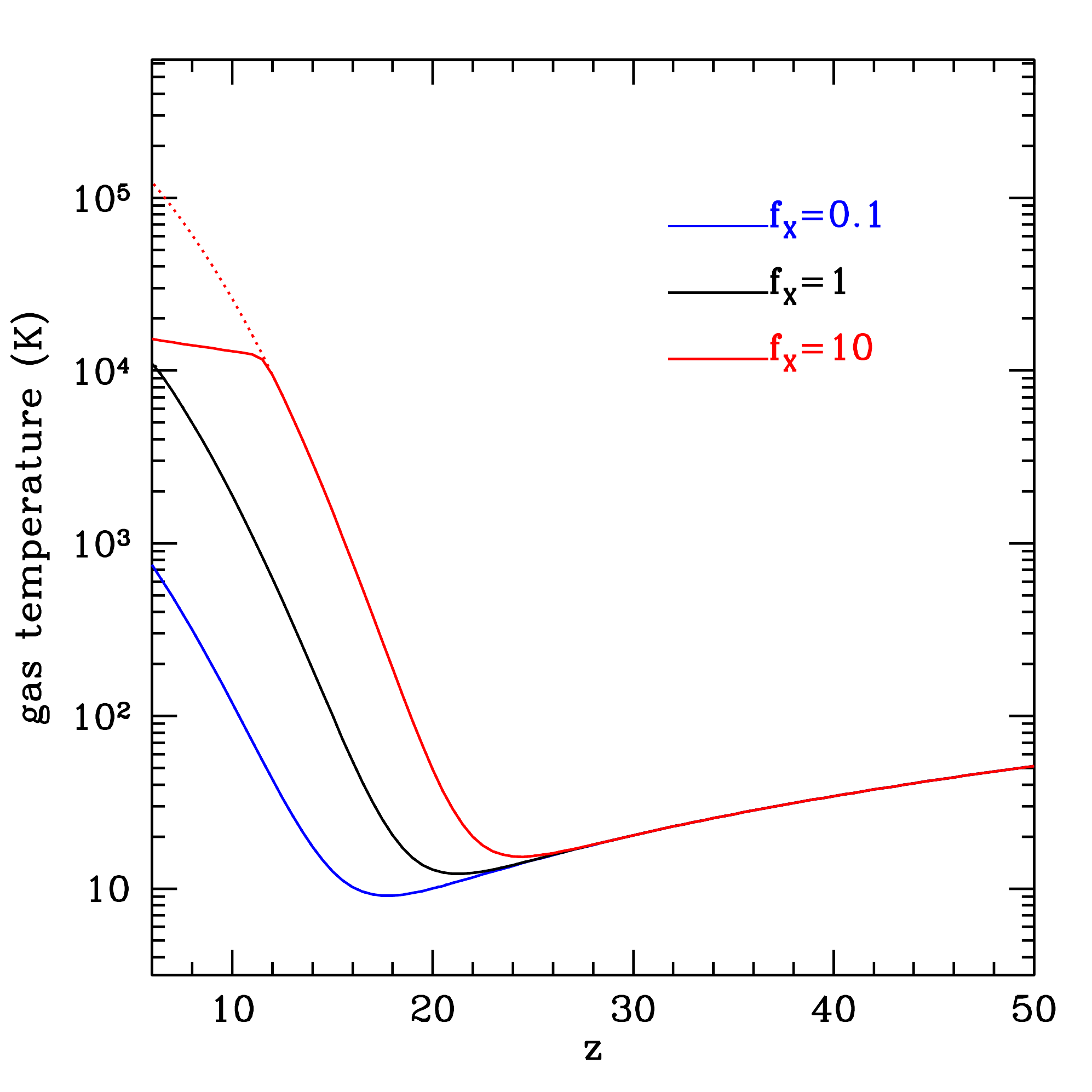}
      \caption{Thermal history of the low-density neutral inter galactic medium (IGM) due to heating by accreting stellar black holes in high mass X-ray binaries (BH-HMXBs). This figure shows the gas temperature of the IGM as a function of redshift z for three possible values of f$_X$  as defined in Eq. (6):  f$_X = 0.1$ [blue dashed line],  f$_X = 1.0$ [red dotted line],  f$_X = 10.0$ [black solid line]). As discussed in the text, most likely f$_X >1$.
              }
         \label{tplot}
   \end{figure}

It is an open question how significant may be boosting the X-ray emissivity of star-forming galaxies in the high-redshift Universe for the global ionized fraction (i.e. averaged over the entire volume of the observable Universe). However, it has been shown  (\cite{Furlanetto06a}) that soft X-rays (E$_{\gamma} < 2$ keV, \cite{ Pritchard07}) and inverse-Compton scattering from relativistic electrons could have profound implications for the amount of heating of the low-density neutral IGM. BH-HMXBs are powerful sources of soft X-rays and relativistic jets, and their heating could in turn affect the overall reionization process indirectly in ways that will need to be investigated further.  In Fig ~\ref{tplot} we show that f$_X$  (as defined in equations 5 and 6) is the parameter that determines the thermal history of the IGM.  Increasing f$_X$ causes the neutral IGM to be heated earlier. As shown in the thermal evolution for f$_X = 10.0$, when the gas temperature approaches $10^4$ K, cooling through collisional excitation of the atomic hydrogen becomes efficient, and further heating is  not possible in practice. As discussed in section 8,  the formation of low-mass galaxies in the neutral IGM at z=10-20 is suppressed when the gas temperature is as high as $10^4$ K, namely, for  f$_X > 5$. For further details on this particular model we refer the reader to Pritchard \& Furlanetto (2007).

\section{The 21cm line of HI during reionization}
\begin{figure}
   \centering
   \includegraphics[width=8.cm]{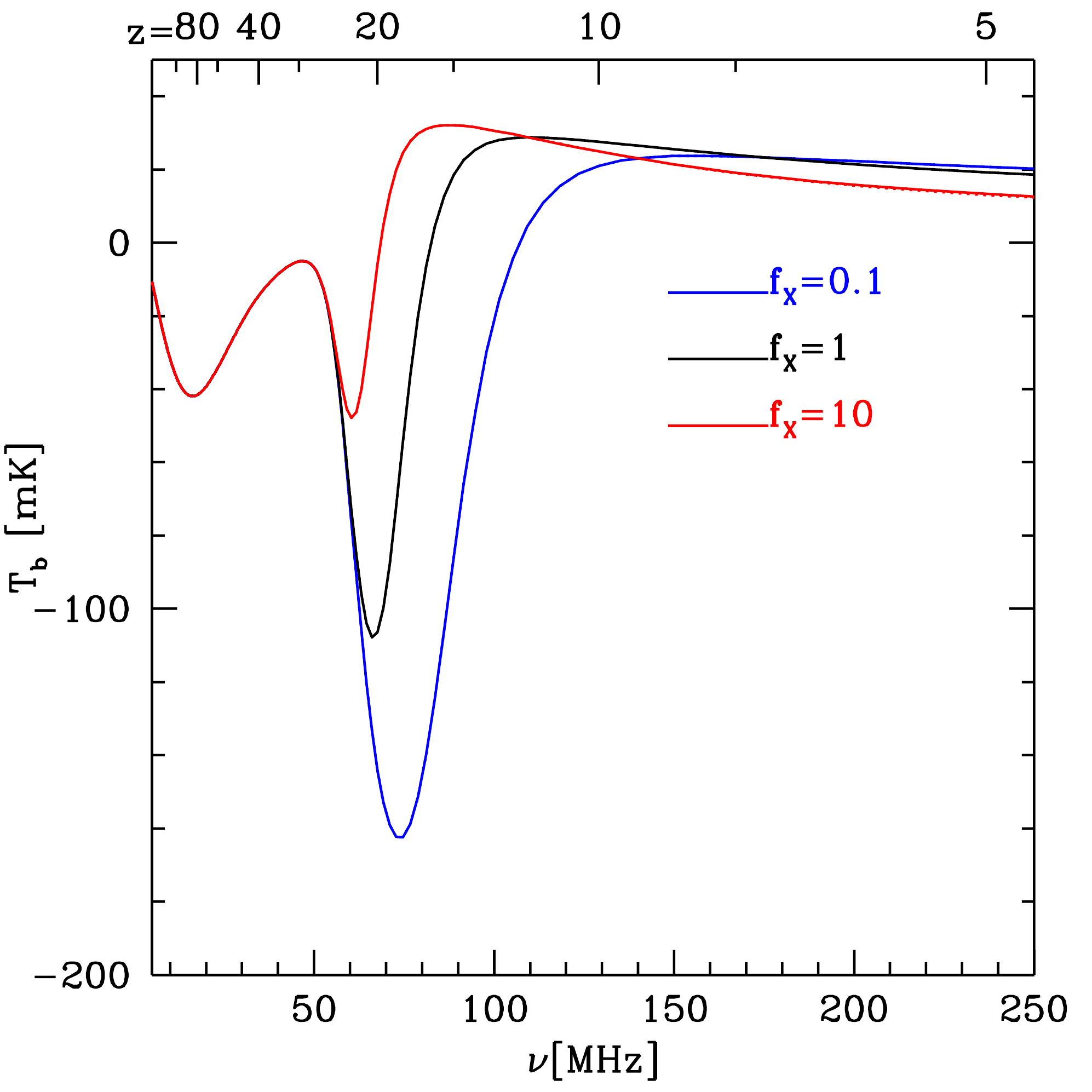}
      \caption{Brightness temperature of the hyperfine transition of the ground state of atomic hydrogen (the wavelength of this transition is 21 cm), averaged over the entire sky, as a function of redshift z for the same three different values of f$_X$ as in Fig.~\ref{tplot}.  
              }
         \label{tplot_nu}
   \end{figure}

The precise temperature evolution of the neutral IGM is known to strongly affect the global 21 cm signature expected from neutral HI during the epoch of reionization (\cite{Furlanetto06b}). In Fig~\ref{tplot_nu}, we plot the evolution of the brightness temperature (averaged over the entire sky) as a function of redshift following the same prescriptions as in Pritchard \& Loeb (2010), again for the same three models with different values of f$_X$. The gas temperature couples to the excitation temperature through collisions, and through scattering of Lyman $\alpha$ photons. When this excitation temperature, also known as the spin temperature, is higher (lower) than the temperature of the cosmic microwave background (CMB), then it is possible to observe hydrogen atoms in emission (absorption) against this CMB. The difference in spin and CMB temperature is referred to as the ``brightness" temperature (\cite{Furlanetto06b}). Clearly, when the gas is heated earlier, the spin temperature can increase earlier, and hydrogen can be seen in emission earlier in the evolution of the Universe. Increasing f$_X$ also reduces the interval in redshift - and therefore - frequency over which hydrogen can be seen in absorption against the CMB (corresponding to negative brightness temperature). Single dipole experiments such as EDGES are currently attempting to measure this signal  (\cite{Bowman08}). One of the present challenges in observational astronomy is to directly observe this 21 cm signal from neutral hydrogen in the young Universe (\cite{Morales10}). This can be accomplished with single radio dipole experiments, such as EDGES (\cite{Bowman08}), which are potentially capable of detecting the global 21 cm signal (\cite{Pritchard10}) at redshifts z $< 30$, directly measuring early heating of the IGM.

A boosted X-ray emissivity of star forming also affects the fluctuations in the 21 cm background (\cite{Pritchard07};    \cite{Pritchard08}). Detecting these fluctuations is one of the prime scientific drivers for the next generations of radio interferometers such as the MWA, LOFAR, and SKA. The fluctuations in the 21 cm background radiation contain the largest amount of cosmological information (\cite{Loeb04}), and are therefore invaluable constraints on cosmological parameters (as well as fundamental physics). However, ``astrophysics" introduces additional fluctuations, for example, through temperature fluctuations and through fluctuations in the Lyman $\alpha$ background (\cite{Pritchard07};  \cite{Pritchard08}). Both are sourced (on different scales) by galaxies that themselves provide biased tracers of the underlying density field. To fully exploit the rich data set provided by the 21 cm fluctuations therefore requires a good understanding of the nature of the astrophysical sources illuminating, heating, and ionizing the hydrogen in our Universe.  Interestingly, one heats the neutral IGM earlier by boosting f$_X$, which implies that temperature fluctuations are suppressed during the later stages of reionization, which improves the prospects for extracting cosmological information from the 21 cm background  (\cite{Pritchard07}; \cite{Pritchard08}).

\section{The role of stellar black holes in the formation of dwarf galaxies}

The cold dark matter model of the universe provides the framework for significant progress in understanding the
large-scale properties and physical principles that govern the large-scale evolution of the universe during the first 400 thousand years.
However, it is still poorly understood how the first stars and black holes in galaxies were formed and the way that in
less than a billon year these pristine objects re-ionized and re-heated most of the matter in
the universe  over large volumes of space. The apparent disparity between the number of dwarf galaxies predicted by the cold dark matter model of the universe
and the number of small galaxies observed so far in the halo of the Galaxy is a subject of topical interest in cosmology (\cite{Loeb10}). Power et al. (2009) had already pointed out the possible implications of X-ray binaries in primordial globular clusters for the reionization, hence, galaxy formation at high redshifts.

It is believed that the first stars had formed in gas clouds with a virial temperature of a few hundred K due to cooling by molecular hydrogen, H$_2$ (\cite{Loeb10}). But the UV radiation produced by these stars could have easily dissociated H$_2$, making atomic hydrogen (H I) cooling necessary for further star formation. The galaxies that reionized the IGM were therefore likely to have a virial temperature above the H I cooling threshold of $10^4$ K.  X-ray, and UV heating by BH-HMXBs of the diffuse IGM during reionization would have resulted in an additional increase in the total minimum galaxy mass M$_{min}$

\begin{equation}
\begin{array}{l}
M_{min}   =
10^9
\left(\frac{\mathrm{\rho}}{\mathrm{100 \rho_C}}\right)^{-\frac{1}{2}}
\left(\frac{\mathrm{\mu}}{\mathrm{0.6}}\right)^{-\frac{3}{2}}\\
\hspace{4.5cm}
\left(\frac{\mathrm{T(K)}}{\mathrm{10^4}}\right)^{\frac{3}{2}}
\left(\frac{\mathrm{1+z}}{\mathrm{10}}\right)^{-\frac{3}{2}}M_{\sun}, 
\end{array}
\end{equation}

\noindent where $\rho_C$ is the critical mass density for a flat Universe, $\rho$ the mass density in the galaxy, $\mu$ the mean molecular weight, z the redshift, and T the temperature of the IGM.
Once the IGM was heated to a temperature of $10^4$ K, dark matter halos with masses below $10^9 M_{\sun}$ could no longer accrete IGM material because the temperature of the infalling gas increased by an extra order of magnitude as its density increased on its way into these galaxies. In that regime, only gaseous halos with virial temperatures above $10^5$ K could have accreted fresh IGM gas and converted it to stars. The census of dwarf galaxy satellites of the Milky Way requires a related suppression in the abundance of low-mass galaxies relative to low-mass dark matter halos (see \cite{Munoz09}  and references therein). The thermal history of the IGM therefore has a direct impact on the properties of the faintest galaxies at high redshifts, as well as on the smallest dwarf galaxies in the local universe.

It is interesting to note that black holes of different mass scales play a role in galaxy formation. Feedback from supermassive black holes halt star formation, quenching the unlimited mass growth of massive galaxies (\cite{Cattaneo09}), and we show here that feedback from stellar black holes in HMXBs during the reionization epoch suppress the number of low-mass dwarf galaxies. Therefore, BH-HMXBs in the early universe are an important ingredient in reconciling the apparent disparity between the observed number of dwarf galaxies in the Galactic halo with the number of low-mass galaxies predicted by the cold dark matter model of the universe.

\section{Conclusions}

The main conclusions of this work are the following:

 \begin{enumerate}
  
 \item The  ratio of black holes to neutron stars and the ratio of black hole binaries to solitary black holes should increase with redshift; that is, the rate of
formation of BH-HMXBs was significantly  higher in the early Universe than at present. 

 \item Feedback from one of those BH-HMXBs during its whole lifetime can be more than $10^{54} erg$ (orders of magnitude larger than that of the photonic and barionic energy from a typical core collapse supernova).

 \item An accreting black hole in a high-mass binary emits a total number of  ionizing photons that is comparable to its progenitor star, but one X-ray photon emitted by an accreting black hole may cause the ionization of several tens of hydrogen atoms in a fully neutral medium. 

 \item The most important effect of BH-HMXBs in the early universe is the heating of the IGM. Soft X-rays and inverse-Compton scattering from relativistic electrons produced by BH-HMXBs  heat the low-density medium over large distances to temperatures of  $\approx 10^4$ K, which limits the recombination rate of hydrogen and keeps the IGM ionized. 

 \item A temperature of the IGM of $\approx 10^4$ K  limits the formation of faint galaxies at high redshifts. It constrains  the total mass of dwarf galaxies to $\geq 10^9 M_{\sun}$ .

 \item BH-HMXBs in the early universe are important ingredients for reconciling the apparent disparity between the observed number of faint dwarf galaxies in the Galactic halo with the number of low-mass galaxies predicted by the cold dark matter model of the universe.

\item An additional effect of metallicity in the formation of BH-HMXBs (\cite{Mirabel10}) is to boost the formation 
of BH-BH binaries as more likely sources of gravitational waves than NS-NS systems (\cite{Belczynski10}).
 
\end{enumerate}

\begin{acknowledgements}
I.F.M thanks the referee G. Meynet, and A. King, P. Fabbiano, T. Piran, T. Linden and V. Kalogera for useful information and kind comments. This work was supported in part by NSF grant AST-0907890 and NASA grants NNX08AL43G and NNA09DB30A.

\end{acknowledgements}

\end{document}